\documentclass[aps,preprint,showkeys, amsfonts,showpacs,amsmath,amssymb]{revtex4-1}
\usepackage[dvips]{color}
\usepackage{epsfig}
\usepackage{amsmath}
\usepackage{graphicx}

\begin{document}
\title{A 5D HOLOGRAPHIC DARK ENERGY IN DGP-BRANE COSMOLOGY}

\author{H. Farajollahi}
\email{hosseinf@guilan.ac.ir}
\affiliation{School of Physics, Unversity of New South Wales, Sydney, NSW,
2052, Australia}
\author{A. Ravanpak}
\email{a.ravanpak@vru.ac.ir}
\affiliation{Department of Physics, Vali-e-Asr University, Rafsanjan, Iran}

%\date{\small {\today}}

\begin{abstract}
This paper is aimed at investigating a 5D holographic dark energy in DGP-BRANE cosmology by employing a combination of Sne Ia, BAO and CMB observational data to fit the cosmological parameters in the model. We describe the dynamic of a FRW for the normal branch ($\epsilon$ = +1) of solutions
of the induced gravity brane-world model. We take the matter in 5D bulk as holographic dark energy that its holographic nature
is reproduced effectively in 4D. The cosmic evolution reveals that the effective 4D holographic dark energy behaves as quintessence while taking into account the 4D cold dark matter results in matter dominated universe followed by late time acceleration.
\end{abstract}
%\pacs{}

\keywords{holography, dark energy, brane, DGP, constraint}
\maketitle

%\newpage

\section{INTRODUCTION}

The exciting and ingenious idea of holographic dark energy that has recently attracted many researchers is capable to interpret current cosmic acceleration. \cite{Li}-\cite{Farajollahi}. The idea initiated from the cosmological application of
the more fundamental holographic principle and despite some objections it reveals the dynamical nature of the vacuum energy by relating it to
cosmological volumes. The holographic principle states that due to the limit set by the formation of a black hole, in effective field theory, the UV Cut-off, $\Lambda_{uv}$, is related to the IR Cut-off $L$ as $L^3\Lambda_{uv}^4\leq L M_p^2$ where $M_p$ is reduced Planck mass
. The effective field theory describes all states of system except those already collapsed to a black hole and the
 vacuum energy density via quantum fluctuation is given by $\rho_{vac}\sim \Lambda_{uv}^4\sim M_p^2 L^{-2}$ where $L$ is characteristic length scale of the universe
. From vacuum energy density the dark energy density caused via quantum fluctuation is given by $\rho_{de}\sim 3d^2 M_p^2 L^{-2}$ where $d$ is a model parameter. By taking different characteristic length scale we can construct various holographic dark energy models.

On the other hand, a large amount of current research heads towards higher dimensional gravity and in particular brane cosmology \cite{Antoniadis},\cite{Randall2},
in which assumes our Universe as a brane embedded in higher dimensional spacetime. A well-known example of brane cosmological model is the Dvali-Gabadadze-Porrati (DGP) braneworld model \cite{Dvali}, in which the 4D world is a FRW brane
embedded in a 5D Minkowski bulk. On the 4D brane the gravity action is proportional to $M_P^2$ whereas in the bulk it is proportional to the corresponding
quantity in 5D, $M_5^3$. The total energy-momentum is confined on a
3D brane embedded in a 5D infinite volume
Minkowski bulk. There are two different ways to confine the 4D brane into the 5D spacetime;
the DGP model has two separate branches denoted by $\epsilon=\pm 1$
with distinct features. The $\epsilon$ = +1 branch is capable to interpret the current cosmic acceleration without any need to introduce dark
energy, whereas for the $\epsilon$ = -1 branch, dark energy is needed \cite{Deffayet},\cite{Chimento}.

Holographic dark energy in the context of DGP brane cosmological models have been investigated in \cite{Wu2} and \cite{Setare3}. But, in both of them a 4D holographic dark energy model has been used. Recently, the extended 5D holographic dark energy model has been considered in \cite{Kim}, but it is grounded on unstable and non-physical arguments. For more studies about this topic see \cite{Saridakis} and \cite{Saridakis2}.

The main motivation of 5D holographic dark energy is that in a 5D spacetime, the natural framework for the cosmological application in connection with dark energy of holographic principle is the space of the bulk and not the 4D brane, because the attributes of gravitational theory or quantum field, such as cut-off's and vacuum energy can be determined by the maximal uncompactified space of the model. For example, in braneworld models, black holes are generally D-dimensional \cite{Kanti},\cite{Cavaglia}, irrespective of their 4D effective effects. Consequently, although the holographic principle is applicable to any dimensions \cite{Bousso},\cite{Iwashita}, its cosmological application on the subject of dark energy must be considered in the maximally-dimensional subspace, i.e. in the bulk. Afterwards, this 5D holographic dark energy results in an effective 4D dark energy component in the Friedmann equation on the brane with inherent
holographic nature.

In this work we present the 5D holographic dark energy in DGP brane cosmology. We will see that under this discussion we can generate an effective 4D holographic dark energy from the Hubble horizon and so some of the holographic dark energy model problems, related to future event horizon as the length scale, like causality problem and circular logic problem are removed, automatically. A simple and not very similar work at this topic has been done in a different way in \cite{Liu}. Here, we will constrain the cosmological parameters in the model, both analytically and also using observational data.

Also, in \cite{Bamba}, the authors have shown the equivalent description of different
theoretical models of dark energy. They have examined the fluid and scalar dark energy descriptions of various models such as holographic dark energy. We should notice that our model follows this organism, as well and we can see this from the numerical results. But at a special case we will show the relation between our model and the $\Lambda$CDM model, when the dark energy dominated regime is considered.

\section{HOLOGRAPHIC DARK ENERGY IN THE BULK}

In this section we review the basic results of a 5D bulk holographic dark energy. For a 5D spherically symmetric and uncharged black hole, its mass $M_{BH}$ is related to its Schwarzschild radius $r_s$ via $M_{BH} = 3\pi M_5^3r_s^{2}/8$ where $M_5$ is the 5D Planck mass and is related to the 5D gravitational
constant $G_5$ with $M_5 = G_5^{-\frac{1}{3}}$. Moreover, the 4-dimensional Planck mass $M_P$ given by $M_P^2 = M_5^{3}V_{1} $ where $V_{1}$ is the volume of the extra-dimensional space. For the holographic dark energy in the 5D bulk we have $\rho_{\Lambda 5}V(S^{3})\leq M_{BH}$ where $\rho_{\Lambda 5}$ is the 5D bulk vacuum energy, and $V(S^{3})$ is the volume of the maximal hypersphere in a 5D spacetime,
given as $V(S^{3}) = \pi^2r^{4}/2$. The dark energy density, $\rho_{\Lambda 5}$, can be viewed as holographic dark energy,
\begin{equation}\label{HDE}
    \rho_{\Lambda 5} = c^2M_{BH}V^{-1}(S^{3})=\frac{3c^2M_5^{3}}{4\pi}L^{-2},
\end{equation}
with $L$ as a large distance and $c^2<1$. Similar to 4-dimensional case, the distance $L$ can be Hubble radius, particle horizon, or the most appropriate future event horizon, for a flat Universe.

Next, we study the holographic dark energy in 5D DGP-brane cosmology with the action
of the form
\begin{eqnarray}\label{action}
S &=& \frac{1}{16\pi}M_5^3\int_{{\cal M}_5}d^5x\sqrt{-^5g}(^5R+{\cal L}_m) \nonumber\\ &+& \frac{1}{16\pi}M_P^2\int_{\sum}d^4x\sqrt{-^4g}(^4R+{\cal K}+{\cal L}_m),
\end{eqnarray}
where the ${\cal M}_5$ and $\sum$ indicate respectively bulk (B) and brane (b) and ${\cal K}$ is the trace of extrinsic curvature. The extra term in the
boundary introduces a a cross-over length scale $r_c = \frac{M_P^2}{2M_5^3}$ which separates two different
regimes of the theory. For distances much smaller than $r_c$ one would expect the solutions to be well approximated
by general relativity and at larger distances the modifications takes into account. In FRW cosmology, the Friedmann equation on the brane is \cite{Phong}
\begin{equation}\label{fried}
H^2 = \frac{8\pi\rho_b}{3M_P^2}-\frac{k}{a^2}+\frac{\epsilon}{r_c}\sqrt{H^2-\frac{4\pi\rho_B}{3M_5^3}
+\frac{k}{a^2}+\frac{\xi}{a^4}}.
\end{equation}
We assume a flat universe, $k = 0$, and a vanishing last term $\xi = 0$. If also consider the matter density in the bulk a holographic dark energy, $\rho_B = \rho_{\Lambda5}$ and the matter on the brane as a cold dark matter $\rho_b = \rho_m$, then we have
\begin{equation}\label{friedmann}
H^2 = \frac{8\pi\rho_m}{3M_P^2}+\frac{\epsilon}{r_c}\sqrt{H^2-\frac{4\pi\rho_{\Lambda5}}{3M_5^3}}.
\end{equation}
Alternatively, we may rewrite the Friedmann equation in the conventional form
\begin{equation}\label{traditional}
H^2 = \frac{8\pi\rho_m}{3M_P^2} + \frac{8\pi\rho_\Lambda}{3M_P^2}
\end{equation}
where the 4D dark energy is
\begin{equation}\label{4DDE}
\rho_\Lambda\equiv\rho_{\Lambda4}=\frac{3M_P^2\epsilon}{8\pi r_c}\sqrt{H^2-\frac{4\pi \rho_{\Lambda5}}{3M_5^3}}.
\end{equation}
From (\ref{HDE}) we find the 5D bulk holographic dark energy, $\rho_{\Lambda5}$, and finally arrive at the following form for
the effective 4D holographic dark energy
\begin{equation}\label{effective4DDE}
   \rho_\Lambda =  \frac{3M_P^2\epsilon}{8\pi r_c}\sqrt{H^2-c^2L^{-2}}.
\end{equation}
In a flat 4D spacetime, the future event horizon is taken as the the most appropriate cut-off scale that fits holographic statistical physics \cite{Li}. Similarly, in \cite{Saridakis} the author shows that in the 5D extension of 4D, still future Event horizon is preferable. In this work, we show that in case of a 5D DGP holographic model of dark energy the Hubble radius can be taken as a cut-off scale that suitably fit the observational data. With the Hubble radius as the horizon, i.e. $L=H^{-1}$ and by using (\ref{traditional}), the dark energy and matter densities then become
\begin{equation}\label{4DDEapp}
    \rho_\Lambda = \frac{3M_P^2\epsilon H}{8\pi r_c}\sqrt{1-c^2}
\end{equation}
\begin{equation}\label{traditionalapp}
    \rho_m = \frac{3M_P^2}{8\pi}(H^2 - \frac{\epsilon H}{r_c}\sqrt{1-c^2})
\end{equation}
From (\ref{4DDEapp}), one immediately finds two new constraints on $c$ and  $\epsilon$, i.e. $0\leq c\leq1$ and for positive $\rho_\Lambda$, $\epsilon$ must be $1$. The second constraint can also be obtained from a thermodynamic method. In terms of density parameter, $\Omega_\Lambda$, equation (\ref{4DDEapp}) can be rewritten as
\begin{equation}\label{Homega}
    H\Omega_\Lambda = \frac{\epsilon\sqrt{1-c^2}}{r_c}.
\end{equation}
By differentiating of the above equation with respect to $t$ we yield
\begin{equation}\label{Hdot}
    \dot H = \frac{-\epsilon\sqrt{1-c^2}\dot{\Omega_\Lambda}}{r_c\Omega_\Lambda^2}.
\end{equation}
On the other hand, the Gibbons-Hawking entropy is proportional to the squared radius of the universe that is taken as the Hubble horizon, i.e.
\begin{equation}\label{GH}
    S \sim A \sim L^2 \sim H^{-2}.
\end{equation}
Since  entropy should increase during expansion phase, we obtain $\dot H \leq0$. From (\ref{Hdot}) and by assuming that $\Omega_\Lambda$ increases with time we again find $\epsilon = +1$. We should notice that the two constraints on $c$ and $\epsilon$ are the characteristics for this model only if Hubble radius is taken as the horizon. By using $\Omega_\Lambda = 1-\Omega_m = 1-\Omega_{m0}H_0^2H^{-2}(1+z)^{3}$ and $\Omega_{r_c} = 1/(4r_c^2H_0^2)$ in (\ref{Homega}), the Friedmann equation becomes
\begin{equation}\label{BFequation}
    E^2 =2\sqrt{\Omega_{r_c}(1-c^2)}E+\Omega_{m0}(1+z)^{3}.
\end{equation}
where we have used the definition $E^2 = H^2/H_0^2$. This equation at $z=0$ yields the constraint
\begin{equation}\label{Friedcons}
    1-\Omega_{m0}=2\sqrt{\Omega_{r_c}(1-c^2)}.
\end{equation}
In the next section we are going to constraint the model parameters $c$, $\Omega_{m0}$ and $\Omega_{r_c}$ with the observational data by employing $\chi^2$ method. However, with attention to constraint (\ref{Friedcons}) we implement this in two different ways.

{\bf A:}

Using (\ref{Friedcons}) we can neglect $c$ and $\Omega_{r_c}$ in favor of $\Omega_{m0}$ in (\ref{BFequation}) and obtain
\begin{equation}\label{1stFried}
    E^2 =(1-\Omega_{m0})E+\Omega_{m0}(1+z)^{3}.
\end{equation}
After best-fitting the value of $\Omega_{m0}$, with attention to constraints on $c$, (i.e., $0\leq c\leq1$) it seems that we can obtain some constraints for $\Omega_{r_c}$. After rewriting (\ref{Friedcons}) as
\begin{equation}\label{rccons}
    \Omega_{r_c}=\frac{(1-\Omega_{m0})^2}{4(1-c^2)}
\end{equation}
we see that for $c=0$ where shows no holographic dark energy we can find for $\Omega_{r_c}$ a lower limit as $\Omega_{r_c}=(1-\Omega_{m0})^2/4$. Also, for $c=1$ where in our model stands for a universe filled with matter (see (\ref{4DDEapp}) and (\ref{traditionalapp})) we find $\Omega_{r_c}\rightarrow\infty$.

{\bf B}:

From (\ref{Friedcons}) we obtain $\Omega_{m0}$ in terms of $c$ and $\Omega_{r_c}$ and after replacing in (\ref{BFequation}) we reach to
\begin{equation}\label{2ndFried}
    E^2 =2\sqrt{\Omega_{r_c}(1-c^2)}E+(1-2\sqrt{\Omega_{r_c}(1-c^2)})(1+z)^{3}.
\end{equation}
Then, we can best-fit our model parameters $c$ and $\Omega_{r_c}$ in such a way that the condition $\Omega_{m0}\geq0$ is satisfied. It is clear from (\ref{Friedcons}) that for any permitted values of $c$ and $\Omega_{r_c}$ in our model, $\Omega_{m0}$ can not be larger than one.

\subsection{EQUIVALENCE WITH $\Lambda$CDM MODEL}
We know that various dark energy models have been proposed to explain late time acceleration of the universe. Some of them are due to modifications in the matter content of the universe. For example, models with some scalar fields such as quintessence, phantom and tachyon and also models with some kinds of fluid with special equation of states such as Chaplygin gas. On the other hand another set of dark energy models is due to modifications in geometry of spacetime such as $f(R)$ and $f(T)$ gravity models. In a recent work \cite{Bamba}, the authors have shown that any dark energy model may be expressed as the
others.

In this work, we have proposed another dark energy model, thus there should be equivalent descriptions of it in any other models of dark energy. In an attempt to show this equivalence, we assume that we are in a dark energy dominated universe. So, we can neglect the first term at the right hand side of (\ref{traditional}) and then replace $\rho_\Lambda$ by (\ref{4DDEapp}). Thus we reach to
\begin{equation}\label{equivalent}
H = \frac{\epsilon}{r_c}\sqrt{1-c^2}.
\end{equation}
We see that at late time, the Hubble parameter in our model reduce to a constant value. So, in dark energy dominated regime our model is parallel to the $\Lambda$CDM model, the simplest dark energy model which can be interpreted both as a modification in geometry of space-time or in matter content of the universe. See, figure (\ref{fig:3}).

\section{OBSERVATIONAL CONSTRAINTS AND COSMOLOGICAL TEST}

There are a variety of observational data to fit the parameters in cosmological models. Here, we use
Sne Ia which consists of 557 data points belonging to the Union sample \cite{Amanullah}, the baryonic acoustic oscillation (BAO) distance ratio and the cosmic microwave background (CMB) radiation. The constraints from a combination of SNe Ia, BAO and CMB can be obtained by minimizing $\chi^2_{SNe}+\chi^2_{BAO}+\chi^2_{CMB}$. We should notice that to best-fit by SNe Ia data, following \cite{Wu} we have performed a marginalization on the present value of a cosmological parameter called distance moduli $\mu_0$.

\begin{figure}[h]
\centering
\includegraphics[width=0.48\textwidth]{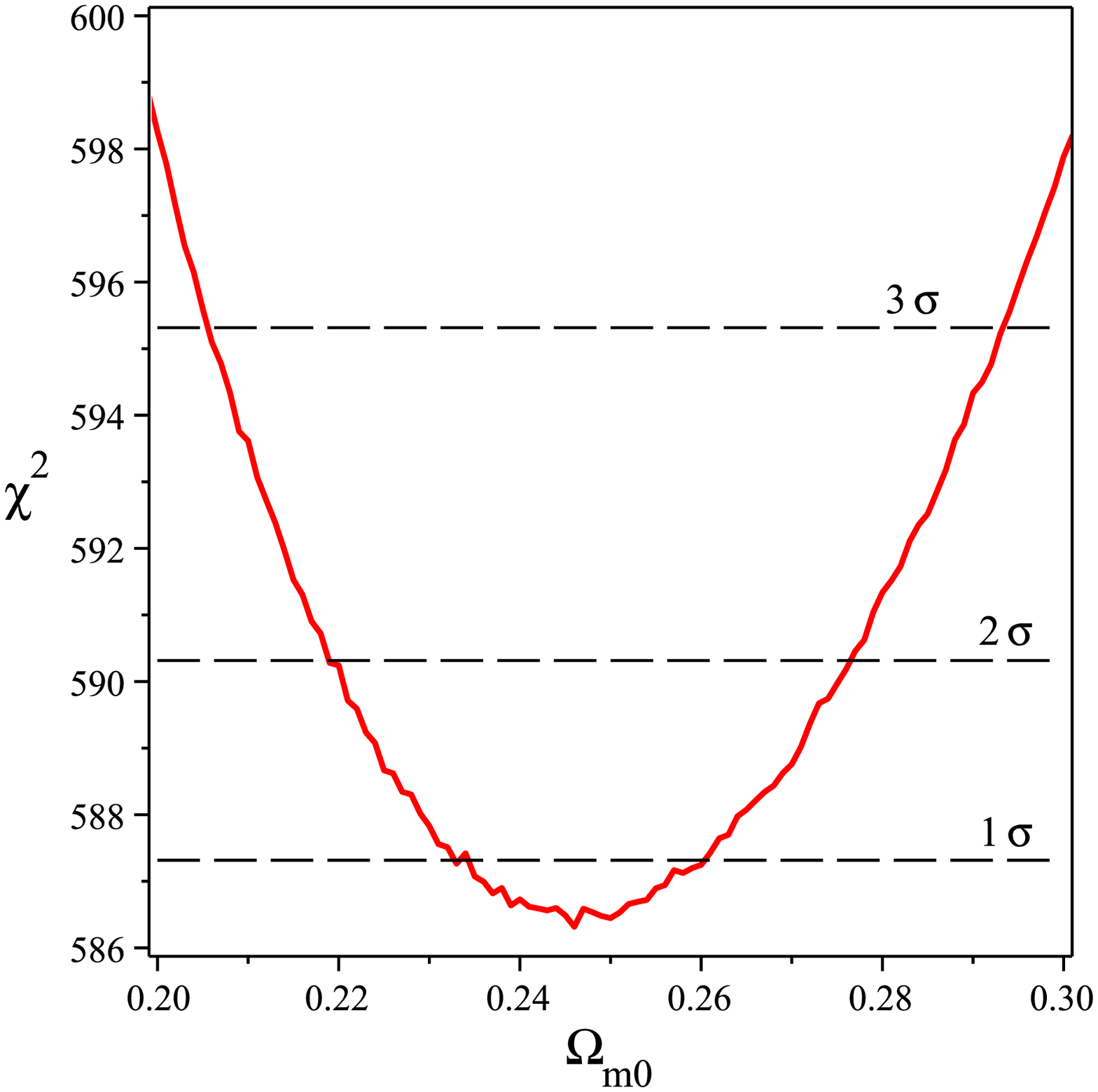}
\includegraphics[width=0.48\textwidth]{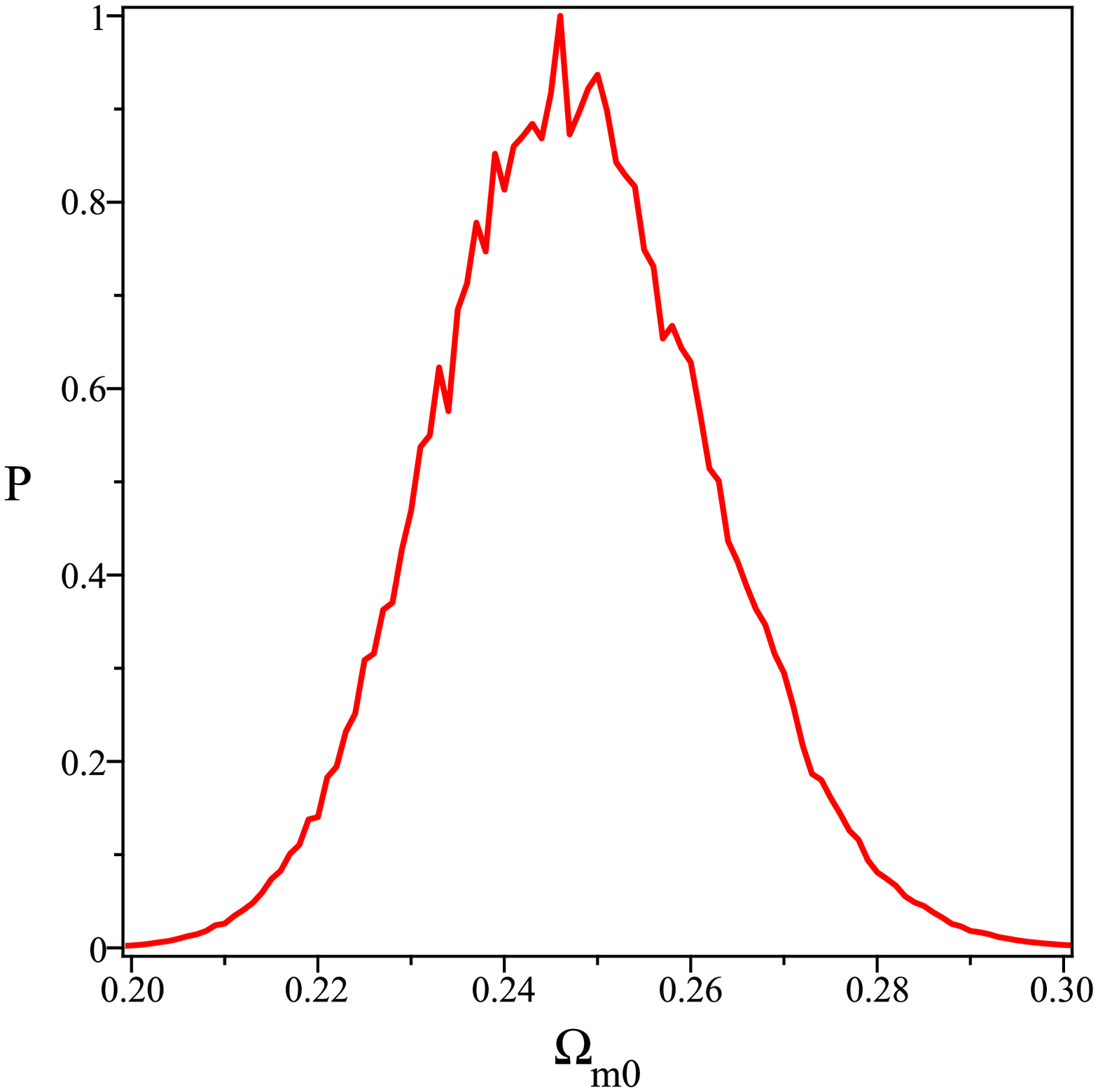}
\caption{left: The evolution of $\chi^2$ with respect to $\Omega_{m0}$. right: The probability distribution with respect to $\Omega_{m0}$.}\label{fig:1}
\end{figure}

For the model, using the first approach we obtain $\Omega_{m0}=0.246^{+0.014 +0.030 +0.047}_{-0.013 -0.027 -0.040}$ for 1$\sigma$, 2$\sigma$ and 3$\sigma$, respectively with $\chi^2_{min}=586.316$. Using (\ref{rccons}) the lower limit for $\Omega_{r_c}$ is obtained as $\Omega_{r_c}\geq0.142$. In figure \ref{fig:1}, left panel we have shown the evolution of $\chi^2$ in terms of $\Omega_{m0}$. The horizontal black dash lines indicate 1$\sigma$, 2$\sigma$ and 3$\sigma$ confidence intervals of $\Omega_{m0}$ where have calculated by adding 1, 4 and 9 to the minimum value of $\chi^2$, respectively. Also, in the right panel we have shown the probability with respect to $\Omega_{m0}$. To this aim we have used the popular definition of probability as P $=\exp(\frac{\chi^2_{min}-\chi^2}{2})$.

The results have been shown in table \ref{table:1}. By $dof$ we mean degrees of freedom, where equals the number of observational data minus the number of parameters we are fitting to data. Note that at the first approach we just have one parameter.

\begin{table}[h]
\caption{bestfit values of the first approach} % title of Table
\centering % used for centering table
\begin{tabular}{|c|c|c|c|c|} % centered columns (5 columns)
\hline\hline %inserts double horizontal lines
observational data  & SNe & SNe + BAO & SNe + CMB & SNe + BAO + CMB \\ %[5ex] % inserts table
%heading
\hline\hline % inserts single horizontal line
$\Omega_{m0}$ & 0.175 & 0.168 & 0.246 & 0.246 \\ % inserting body of the table
\hline % inserts single horizontal line
$\chi^2_{min}$ & 543.270 & 544.456 & 584.383 & 586.316 \\ % inserting body of the table
\hline
$\chi^2_{min}/dof$ & 0.977 & 0.976 & 1.049 & 1.049 \\ % inserting body of the table
\hline
\end{tabular}
\label{table:1} % is used to refer this table in the text
\end{table}\

At the second approach we obtain the best-fit values $c=0.318$ and $\Omega_{r_c}=0.158$ with $\chi^2_{min}=586.287$. Then, using (\ref{Friedcons}) we can calculate $\Omega_{m0}=0.246$ where is exactly the same as the result of the first approach. The results have been shown in table \ref{table:2}.

\begin{table}[h]
\caption{bestfit values of the second approach} % title of Table
\centering % used for centering table
\begin{tabular}{|c|c|c|c|c|} % centered columns (5 columns)
\hline\hline %inserts double horizontal lines
observational data  & SNe & SNe + BAO & SNe + CMB & SNe + BAO + CMB \\ %[5ex] % inserts table
%heading
\hline\hline % inserts single horizontal line
c & 0.884 & 0.382 & 0.786 & 0.318 \\ % inserting body of the table
\hline % inserts single horizontal line
$\Omega_{r_c}$ & 0.778 & 0.202 & 0.370 & 0.158 \\ % inserting body of the table
\hline % inserts single horizontal line
$\chi^2_{min}$ & 543.110 & 544.300 & 584.346 & 586.287 \\ % inserting body of the table
\hline
$\chi^2_{min}/dof$ & 0.979 & 0.977 & 1.051 & 1.051 \\ % inserting body of the table
\hline
\end{tabular}
\label{table:2} % is used to refer this table in the text
\end{table}\

\begin{figure}[h]
\centering
\includegraphics[width=0.48\textwidth]{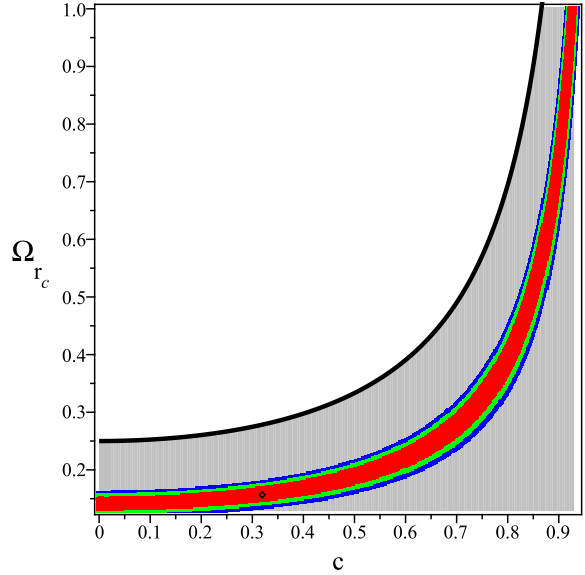}
\caption{The contour plot for 1$\sigma$(red), 2$\sigma$(green) and 3$\sigma$(blue) confidence regions using best-fitted parameters by SNe Ia, BAO and CMB combination. The black circle indicates the best fit point. We have removed the $\Omega_{m0}<0$ region (white area) by the black curve that shows $\Omega_{m0}=0$.}\label{fig:2}
\end{figure}

Fig. (\ref{fig:2}) shows the confidence contours of the parameters $\Omega_{r_c}$ and $c$ for combination of SNe Ia, BAO and CMB. The black curve and the gray area show $\Omega_{m0}=0$ and $\Omega_{m0}>0$, respectively.

\begin{figure}
\centering
\includegraphics[width=0.48\textwidth]{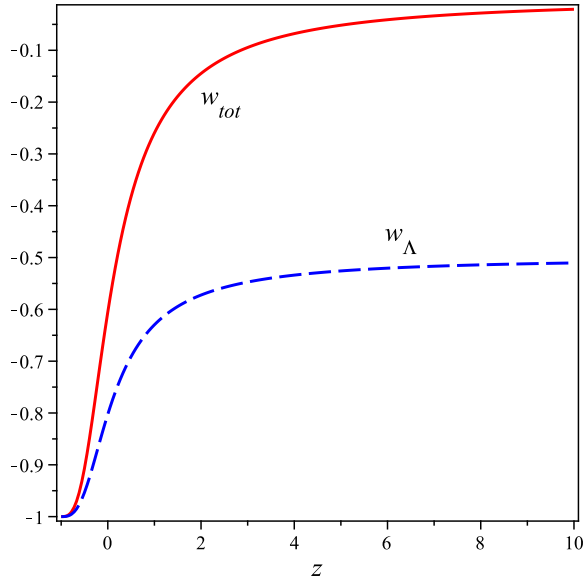}
\caption{The evolutionary curve of DE and total EoS parameter with respect to redshift.}\label{fig:3}
\end{figure}

Using the best-fitted constrained model parameters, figure (\ref{fig:3}), left panel, shows the evolutionary curve of the holographic dark energy EoS parameter for the best fitted values of our model. One can see that the parameter becomes tangent to $-1$ in the future. If we assume that the holographic dark energy is conserved on the brane as
\begin{equation}\label{conservation}
    \dot\rho_\Lambda + 3H(1+w_\Lambda)\rho_\Lambda = 0,
\end{equation}
then using (\ref{4DDEapp}) and the definition of $E$ we obtain
\begin{equation}\label{eos}
w_\Lambda = -1+\frac{(1+z)}{3E}\frac{dE}{dz}\cdot
\end{equation}

Also, we can express the total EoS parameter generally as
\begin{equation}\label{eostot}
    w_{tot} = -1+\frac{2(1+z)}{3E}\frac{dE}{dz}\cdot
\end{equation}
In figure (\ref{fig:3}), the evolutionary curve of DE and total EoS parameter for the best fitted values has been shown.

At the end of this section we have a comparison with \cite{Wu2}. As we mentioned in introduction, at this work the authors have used a 4D holographic dark energy model in the DGP brane cosmology. They have investigated all kinds of length scale, i.e. Hubble horizon, particle and future event horizon. They have shown that if the Hubble horizon is considered as the IR cut-off, in $\epsilon=+1$ branch, the best fit values from SNe Ia+BAO data, prefer a pure DGP model with negligible vacuum energy. But in our article, we have shown that in a 5D holographic dark energy in DGP brane cosmology, choosing Hubble horizon as the length scale causes an effective 4D holographic dark energy which in spite of describing late time acceleration of the universe, removes important problems of an ordinary 4D holographic dark energy model, such as causality and circular logic problems.

\section{SUMMARY AND REMARKS}

We proposed a 5D holographic dark energy scenario in DGP-Brane cosmology. We fitted the model parameters in two different ways by using Sne Ia, Sne Ia+BAO and Sne Ia+CMB and Sne Ia+BAO+CMB dataset. It is noteworthy that the model fits different combination of dataset equally well with the $\chi^2_{min}\rightarrow 1$. However, the combination of Sne Ia+BAO+CMB dataset constrain $\Omega_{m0}$ slightly better.

The fitting model strongly suggest that the accelerated expansion is caused by the 5D holographic dark energy which transferred to an effective 4D dark energy. Without CDM in the 4D matter, i.e., considering a dark energy dominated Universe we found that our model is equivalent with the $\Lambda$CDM model. Also, here we considered Hubble radius as the length scale where caused some special constraints on our model parameters, $c$ and $\epsilon$.

\end{document}